\documentclass[12pt]{article}
\usepackage{a41}
\usepackage{epsfig}
\usepackage{axodraw}
\usepackage{eucal}

\newcommand{\gsim}{\stackrel{>}{\sim}}

\newcommand\GeV{\,\mbox{GeV}}

\begin{document}
\setlength{\baselineskip}{0.515cm}
\sloppy
\thispagestyle{empty}
\begin{flushleft}
DESY 01--007 \hfill
{\tt hep-ph/0102025}\\
January  2001 \\
\end{flushleft}

\setcounter{page}{0}

\mbox{}
\vspace*{\fill}
\begin{center}{
\LARGE\bf Twist--4 Gluon Recombination Corrections}

\vspace{2mm}
{\LARGE\bf for Deep Inelastic Structure Functions}$^{^{\footnotemark}}
$\footnotetext{Work supported in part by EU contract FMRX-CT98-0194
(DG 12 - MIHT) and the National Natural Science Foundation of China
(No. 10075020)}

\vspace{5em}
\large
J. Bl\"umlein$^a$, V. Ravindran$^{a,b}$, Jianhong Ruan$^c$ and 
Wei Zhu$^{a,c}$
\\
\vspace{5em}
\normalsize
{\it $^a$DESY--Zeuthen}\\
{\it Platanenallee 6, D--15735 Zeuthen, Germany}\\

\vspace{1mm}
{\it $^b$Harish--Chandra Research Institute, 
}\\
{\it Chhatnag Road, Jhusi, Allahabad, 211019, India}\\

\vspace{1mm}
{\it $^c$Department of Physics, East China Normal University,}\\
{\it Shanghai 200062, PR China}\\
\vspace*{\fill}
\end{center}
\begin{abstract}
\noindent
We calculate twist--4 coefficient functions for the deep inelastic 
structure function $F_2(x,Q^2)$ associated to 4--gluon operator matrix 
elements for general values of the Bjorken variable $x$ and study the
numerical effect on the slope $\partial F_2(x,Q^2)/\partial \log Q^2$. 
It is shown that these contributions diminish the strongly rising 
twist--2 terms towards small values of $x$.
\end{abstract}

\vspace{1mm}
\noindent
\begin{center}
PACS numbers~:~12.38.Aw, Bx, Qk; 13.60.-r
\end{center}

\vspace*{\fill}
\newpage
\section{Introduction}

\vspace{1mm}
\noindent
The strong rise of deeply inelastic structure functions $F_2(x,Q^2)$ and
$F_L(x,Q^2)$ for small values of the Bjorken scaling variable $x = Q^2/
2 p.q$ is a consequence of both the structure of the twist--2 $(\tau = 2)$
evolution
kernels at leading order and the behavior of the non--perturbative 
flavor singlet input distributions at scales $Q^2_0 \sim 1 \GeV^2$.
Potentially this behavior violates unitarity which has to be guaranteed
by higher order QCD corrections. As well--known, already the
next-to-leading order corrections diminish the rising rate, see 
e.g.~\cite{WKT}\footnote{Recent numerical estimates for NNL 
order~\cite{NV} continue this trend.}, however, they do not lead to a 
saturation of $F_{2,L}(x,Q^2)$ as $x \rightarrow 0$.

The large growth of $F_2(x,Q^2)$ in the double--logarithmic approximation
with idealized \mbox{$\delta(1-x)$-}like 
non--perturbative input distributions
led Gribov, Levin and Ryskin~\cite{GLR} to propose a non--linear 
integro--differential equation for the gluon density, which obeyed
saturation as $x \rightarrow 0$. In this scenario it is assumed that the
non--perturbative $n$--particle gluon distribution is the $n$--th power
of the single--particle gluon density. The evolution kernels are derived
using the double--logarithmic approximation and an idealized 
factorization between the non--perturbative input densities and the
evolution kernels is assumed. The negative sign of the non--linear 
gluon--recombination term is introduced referring to the 
Abramowskii--Gribov--Kancheli cutting rules~\cite{AGK} from Regge theory.
The resummation equation can be expanded into an infinite series of 
so--called fan-diagrams of $O(1/(Q^2)^k)$. This equation is of the 
Glauber--type~\cite{GLAUM} and not derived as a resummation of certain
dominant classes of contributions emerging in the light cone 
expansion, although the term {\it twist} is used by some authors to 
denote the different $1/(Q^2)^k$ contributions.\footnote{More 
appropriately one should call these terms power--corrections.}

Later Mueller and Qiu~\cite{MQ} studied this recombination process too
and derived a value for the recombination strength being yet unspecified
in Ref.~\cite{GLR}. Durand~\cite{DUR} tried to reverse the mechanism
leading to the leading order evolution equations, which, however, turned
out to be not possible at this level.\footnote{A similar mechanism is,
however, applicable in the case of 2--nucleon reactions in a nucleus,
\cite{CRQ}.}
A first numerical illustration of the contributions due the resummation 
in \cite{GLR} on the gluon density was pursued in Ref.~\cite{KWI}. More
detailed investigations, partly leaving the double--log level, were 
carried out in 
Refs.~[10--12].  
The non--linear resummation equation was solved in three different ways 
in \cite{BBS} by showing existence and uniqueness of the solution of
this equation. An exact value of the resummed gluon density in the
saturation limit was derived analytically as well as the solution in the 
quasi-classic limit was studied leading to a derivation of a critical
saturation line as a function of $x$ and $Q^2$.

In a forthcoming investigation of the resummation \cite{GLR} by Bartels
\cite{BAR}, see also \cite{LRS}, it was shown that even using the 
double--logarithmic approximation other than the fan--diagram 
contributions were present in the same order though suppressed 
by a power the number of colors, $N_c$.
Since
the resummation is non--linear these perturbing terms being present in 
all orders spoil the resummation already in this approximation and 
quantitative predictions on their effect are hard to make.

All these investigations have to be viewed in the general context of
perturbative QCD. Here the question arises to which end the assumptions
made can be validated in a rigorous approach. As well--known, the
double--logarithmic approximation often yields much larger results than
the complete calculation, cf.~\cite{BV1}. In a non--linear iteration
quantitative effects due to this difference can be very large. To
separate the non--perturbative input distributions and the perturbative
evolution kernels and coefficient functions factorization has to be
shown for the respective orders in $1/(Q^2)^k$. In general a 
non--universal structure emerges both at the color and Lorentz-level
comparing different orders in $1/Q^2$ and the possibility to resum in
general appears to be unlikely. 

Therefore, in a first step, the simpler question is asked for the 
structure of the 4 gluon $\rightarrow$ 2 quark coefficient functions,
which relates the 4--gluon operator matrix elements to the deep inelastic
structure functions, cf. also Ref.~\cite{BB}. Here a central question is 
that for the sign of these coefficient functions as a function of $x$, 
which in a sense was conjectured to be negative in the foregoing 
literature in the small $x$ limit.
It is clear that one neither can rely on the double--logarithmic 
approximation nor a small-$x$ approximation in the sense of `dominant 
poles', as well--known, see~Refs.~\cite{SUBL1,BV1,SUBL2},
due to the interplay of small $x$ and large $x$ contributions via the
Mellin-type integrals between the coefficient functions and the respective
operator matrix elements. Instead one has to perform a calculation for
the whole range of $x$.

In this note we present numerical results on the effect of main 
contributions to the twist--4 coefficient function driven by a 
two--particle gluon distribution $G_2$ for the slope
$\partial F_2(x,Q^2)/\partial \log Q^2$. This quantity is of interest
since one may generally expect that higher twist effects are more easily 
detectable in this distribution than in the integrated structure 
function
$F_2(x,Q^2)$. After discussing the structure of the non--perturbative
distribution $G_2$ in section~2 the coefficient function is derived
(section 3). Numerical results are presented in section~4 and section~5
contains the conclusions.

\section{4--Gluon Operator Matrix Elements}

\vspace{1mm}
\noindent
The non--perturbative input distributions for a 4--gluon state being
linked to a quark box contains four color, four vector indices, and three
scaling variables  in general. A priori nothing is known about this 
quantity
except the weight factor $Q_0^2/Q^2$ relative to the twist 2 
             contributions defined by the non-perturbative parameter
$Q_0^2 = 1/R^2$. Here $R$ denotes a twist--4 screening length.
In analogy to the single particle gluon 
density one might define the $x$--dependent part for a multi-gluon 
density depending on $n$ momentum fractions
\begin{equation}
G_{(n)}(x_1,\ldots,x_n) = A_n x_1^{\alpha_1} \ldots x_n^{\alpha_n}
(1 - x_1 - \ldots - x_n)^\beta
\end{equation}
similar to Ref.~\cite{KW}. One may further assume 
$\alpha \equiv \alpha_i, \forall i$ for simplicity. If one attempts to 
relate the four--gluon distribution
\begin{equation}
\hat{G}_2(x_1,x_2;x_1',x_2') =
    {G}_2(x_1,x_2;x_1',x_2') \delta(x_1 + x_2  - x_1' - x_2')
\end{equation}
to a product of two single--gluon distributions $G_1(x_i)$ 
it becomes clear
that this is only possible by discarding one of the
${x_i}'s$ or relating it to the other two. In a first attempt one still 
might want to work in such an approximation and even assume that $G_2$ is
found as the simple product of two single gluon densities setting the 
unknown non-perturbative correlator $\chi(x_1,x_2,x_3)$ to $1$ to obtain 
first numerical results. 

It is evident that on the level of twist--4 the general description 
requires the introduction of several 4--particle gluon distributions, 
according to the different color and Lorentz structures, which are not 
related. This further complicates the determination of these quantities
from experimental data. The permutation behavior of the arguments of 
$G_2$ being linked to a certain piece of the coefficient function depends
on the transformation properties of the latter concerning both color and 
the momentum fractions $x_i$. Given the fact that the correlators $\chi$ 
are yet unknown, it is not even clear a priori, whether 
$G_2(x_1,x_2,x_3)$ is positive everywhere. With all these reservations
in mind we are going to follow a {\it simplified} approach in the present
paper. We will
assume $x_1 = x_1' = x_2 = x_2'$ and $\chi \equiv 1$.
This leads to 
\begin{equation}
G_2(x_1) = \frac{Q_0^2}{Q^2} G_1^2(x_1)
\label{ans}
\end{equation}
which guarantees the positivity of $G_2$. Since the ansatz (\ref{ans}) is
not of general validity one has to restrict the investigation using 
special color and Lorentz contractions in the factorization of the 
coefficient function and the hadronic matrix elements below. Following
\cite{MQ} we use for the representation of the
    two--particle (momentum) gluon density
\begin{eqnarray}
{\bf G_2}(x_1) = \frac{9}{8} Q_0^2 x_1^2 G^2_1(x_1)~.
\label{gluf}
\end{eqnarray}
\section{Coefficient Function}

\vspace{1mm}
\noindent
The twist-4 coefficient function describing the 4-gluon $\rightarrow$ 
2-quark amplitude is calculated to $dp_{\perp}^2/p_{\perp}^4$ accuracy 
using time--ordered perturbation theory~\cite{TOPT}. A first investigation
of the problem has been performed in \cite{WZ1,WZ2}.
The contributing 
terms are illustrated in Figure~1 (direct terms) and Figure~2 
(interference terms). Here the grey ovals symbolize the set of diagrams
being shown in Figure~3. The transverse momentum %
$p_\perp$
occurs
in the internal loop.

\vspace*{-1.5cm}
\begin{center}
\begin{picture}(-50,100)(0,0)
\setlength{\unitlength}{0.2mm}
\SetWidth{1.5}
\DashLine(-105,-45)(-105,45){2}
\DashLine(-40,-45)(-40,45){2}
\DashLine(-20,-45)(-20,45){2}
\DashLine(45,-45)(45,45){2}
\Gluon(-130,30)(-80,30){5}{8}
\Gluon(-130,-30)(-80,-30){5}{8}
\Line(-80,30)(-30,30)
\Line(-80,-30)(-30,-30)
\Line(-30,30)(20,30)
\Line(-30,-30)(20,-30)
\CCirc(-30,30){5}{Black}{White}
\GOval(-80,0)(30,8)(0){.45}
\GOval(20,0)(30,8)(0){.45}
\GOval(-130,0)(35,10)(0){1}
\Gluon(20,30)(70,30){5}{8}
\Gluon(20,-30)(70,-30){5}{8}
\GOval(70,0)(35,10)(0){1}
\setlength{\unitlength}{1pt}
\Text(-105,60)[]{$t \rightarrow$}
\Text(-120,50)[]{$x_1$}
\Text(-120,-50)[]{$x_2$}
\Text(60,50)[]{$x_1'$}
\Text(60,-50)[]{$x_2'$}
\end{picture}
\end{center}

\vspace*{2cm} \noindent
\small
{\sf Figure~1:~Direct diagrams contributing to (\ref{coef}). The grey 
oval symbolizes the set of diagrams in Figure~3. Orthogonal dashed lines
stand for the time ordering. The separated white ovals symbolize the
two parts of the non-perturbative 4--gluon  density. $x_1, x_2, x_1'$
and $x_2'$ are the longitudinal momentum fractions, with 
$x_1+x_2-x_1'-x_2' = 0$. The circle stands for the forward
subprocess $\gamma^* + q \rightarrow \gamma^* + q$ through which the
virtual photon couples to the amplitude.}
\normalsize
%
%

\vspace*{-1cm}
\begin{center}
\begin{picture}(100,100)(0,0)
\SetWidth{1.5}
\setlength{\unitlength}{0.2mm}
\DashLine(-105,-45)(-105,45){2}
\DashLine(-43,-45)(-43,45){2}
\DashLine(-70,-45)(-70,45){2}
\DashLine(-5,-45)(-5,45){2}
\Gluon(-130,0)(-80,0){5}{8}
\Line(-80,0)(-30,30)
\Line(-80,0)(-30,-30)
\CCirc(-55,15){5}{Black}{White}
\GOval(-130,0)(35,10)(0){1}
\Gluon(-30,30)(20,30){5}{8}
\Gluon(-30,-30)(20,-30){5}{8}
\Gluon(-30,0)(20,0){5}{8}
\GOval(-30,15)(16,5)(0){.45}
\GOval(-30,-15)(16,5)(0){.45}
\GOval(20,0)(50,10)(0){1}
\DashLine(95,-45)(95,45){2}
\DashLine(157,-45)(157,45){2}
\DashLine(130,-45)(130,45){2}
\DashLine(195,-45)(195,45){2}
\Gluon(70,30)(120,30){5}{8}
\Gluon(70,-30)(120,-30){5}{8}
\Gluon(70,0)(120,0){5}{8}
\GOval(70,0)(50,10)(0){1}
\Gluon(170,0)(220,0){5}{8}
\Line(120,30)(170,0)
\Line(120,-30)(170,0)
\CCirc(145,15){5}{Black}{White}
\GOval(120,15)(16,5)(0){.45}
\GOval(120,-15)(16,5)(0){.45}
\GOval(220,0)(35,10)(0){1}
\setlength{\unitlength}{1pt}
\Text(-105,60)[]{$t \rightarrow$}
\Text(45,0)[]{$+$}
\Text(95,60)[]{$t \rightarrow$}
\end{picture}
\end{center}
\vspace{1cm}
\noindent
\begin{center}
\small

\vspace*{5mm}
{\sf Figure~2:~
~Interference diagrams associated to the process in
Figure~1.}
\normalsize
\end{center}
\normalsize
The white ovals in Figures~1,2 stand for the respective parts of the
non-perturbative distribution ${\bf G_2}$, which are collinearly 
factorized. In the calculation we assumed
$x_1 = x_1'$ and $x_2 = x_2'$, which is the natural choice, if one aims
on using the ansatz ${\bf G_2}(x_1,x_2) = x_1 G_1(x_1) x_2 G_1(x_2)$.
Finally $x_1$ and $x_2$ were, furthermore, set equal, in a first
approximation, according to the choice Eq.~(\ref{gluf}) for ${\bf G_2}$.

\vspace*{-1cm}
\begin{center}
\begin{picture}(220,100)(0,0)
\setlength{\unitlength}{0.2mm}
\SetWidth{1.5}
\Gluon(-130,30)(-80,30){5}{8}
\Gluon(-130,-30)(-80,-30){5}{8}
\Line(-80,30)(-30,30)
\Line(-80,-30)(-30,-30)
\GOval(-80,0)(35,10)(0){.45}
\Gluon(0,30)(50,30){5}{8}
\Gluon(0,-30)(50,-30){5}{8}
\ArrowLine(50,-30)(50,30)
\ArrowLine(50,30)(100,30)
\ArrowLine(100,-30)(50,-30)
\Gluon(130,30)(180,-30){5}{8}
\Gluon(130,-30)(180,30){5}{8}
\ArrowLine(180,-30)(180,30)
\ArrowLine(180,30)(230,30)
\ArrowLine(230,-30)(180,-30)
\Gluon(260,30)(280,0){5}{5}
\Gluon(260,-30)(280,0){5}{5}
\Gluon(280,0)(330,0){5}{8}
\ArrowLine(330,0)(350,30)
\ArrowLine(330,0)(350,-30)
\setlength{\unitlength}{1pt}
\Text(-15,0)[]{$=$}
\Text(115,0)[]{$+$}
\Text(245,0)[]{$+$}
\end{picture}

\vspace*{1.5cm}
\small
{\sf Figure~3: Diagrams symbolized by grey ovals in Figure 1 and 2.}
\normalsize
\end{center}
We have independently recalculated the coefficient functions for these
processes. For the $2 \rightarrow 2$ coefficient function
one obtains~:
\begin{eqnarray}
C^{4 \rightarrow 2;2 \rightarrow 2}_{G \rightarrow q}
(x_1,x_1; x_1',x_1';x)
=
C^{4 \rightarrow 2;2 \rightarrow 2}_{G \rightarrow q}(x_1,x) =
\frac{1}{96} \frac{(2x_1-x)^2}{x_1^5}             [
14 x^2 - 3 x x_1 + 18 x_1^2]~.
\label{coef}
\end{eqnarray}
Under the above assumptions the 3rd diagram in Figure~3 does not 
contribute. Eq.~(\ref{coef}) can be given a simple form introducing the
variable $z=x/x_1$. Note that the Mellin poles in this variable are 
situated at $N=0,-1,\ldots,-4$ and are of similar strength. 
Non of these poles is therefore dominant. For the direct process the
kinematic range of the variable $x_1$ is $x_1~\epsilon~[x/2, 1/2]$.

The interference contributions, Figure~2, consist out of the coefficient
function for the $1 \rightarrow 3$ and $3 \rightarrow 1$ process and
the factorized non-perturbative distributions. One may show, cf. 
\cite{WZ1}, that these coefficient functions are related to that of the 
$2 \rightarrow 2$ process by a sign-change in the {\it whole} $x$ range.
\begin{eqnarray}
C^{4 \rightarrow 2;2 \rightarrow 2}_{G \rightarrow q}
(x_1,x_1; x_1',x_1';x) =
- C^{4 \rightarrow 2;1 \rightarrow 3}_{G \rightarrow q}
(x_1,x_1; x_1',x_1';x) =
- C^{4 \rightarrow 2;3 \rightarrow 1}_{G \rightarrow q}
(x_1,x_1; x_1',x_1';x)~.
\label{coef1}
\end{eqnarray}
However, the kinematic
range of $x_1$ is now $x_1~\epsilon~[x,1/2]$. Arguments were given in 
Ref.~\cite{JAF} that one may choose the same non--perturbative input
for the direct and interference terms. This hypothesis is being adopted
here.
For the slope $\partial F_2(x,Q^2)/ \partial \log Q^2$ one finally
obtains~:
\begin{eqnarray}
\frac{\partial F_2(x,Q^2)}{\partial \log (Q^2)}
&=&
\frac{\partial F_2^{\tau = 2}(x,Q^2)}{\partial \log (Q^2)}
+ \left(\frac{\alpha_s}{2\pi}\right)^2 \frac{1}{Q^2} \int_{x/2}^{1/2}
d x_1 \left(\frac{x}{x_1}\right)
C^{4 \rightarrow 2;2 \rightarrow 2}_{G \rightarrow q}(x_1,x)
{\bf G_2}(x_1) \nonumber\\
& &~~~~~~~~~~~~~~~~
- 2 \left(\frac{\alpha_s}{2\pi}\right)^2 \frac{1}{Q^2} \int_{x}^{1/2}
d x_1 \left(\frac{x}{x_1}\right)
C^{4 \rightarrow 2;2 \rightarrow 2}_{G \rightarrow q}(x_1,x)
{\bf G_2}(x_1)~,
\label{evol}
\end{eqnarray}
where
\begin{eqnarray}
\frac{\partial F_2^{\tau = 2}(x,Q^2)}{\partial \log (Q^2)}
&=&
 \left(\frac{\alpha_s}{2\pi}\right) x \left\{
\sum_{q} e_q^2 \left[
P_{qq} \otimes (q+\overline{q})\right](x) +  [\sum_q e_q^2] \left[
P_{qG} \otimes G_1\right
](x)\right\},
\label{evol1}
\end{eqnarray}
and $\otimes$ denotes the Mellin convolution.
\section{Numerical Results}

\vspace{1mm}
\noindent
Numerical values for the slope $\partial F_2(x,Q^2)/\partial \log Q^2$
are given in Figure~4. Here we compare the results in leading order QCD
for the twist--2 contribution to those obtained including the twist--4
term Eq.~(\ref{evol}). For the twist--2 contribution and the 
non--perturbative distribution ${\bf G_2}(z,Q^2)$ we refer to the
parameterization~\cite{GRV98} and Eq.~(\ref{gluf}), respectively.
Figure~4a and b depict the results for $Q^2 = 5 \GeV^2$ and $10 \GeV^2$,
choosing different screening lengths $R$. Under the above assumptions
the twist--4 correction is negative in the small--$x$ range, i.e.
the screening term in Eq.~(\ref{evol}) wins against the anti--screening
term. The latter one leads to a very small positive contribution only
at large values of $x$. The effect shrinks with rising values of $Q^2$,
and becomes visible only for $x$ values below $x \sim 10^{-4}$ at
perturbative scales of $Q^2 \gsim 4 \GeV^2$. Therefore this effect cannot
be seen yet in the kinematic range of HERA.
Although one might find
arguments for the size of the screening parameter $R$, it is a 
non--perturbative quantity for which a general prediction within QCD
is still missing. We gave illustrations for $R^2 = 5$ and $2 \GeV^2$.
In the latter case a clear effect is visible for $x \leq 2 \cdot 10^{-6}$
and $Q^2 \sim  5 \GeV^2$.

The coefficient function Eqs.~(\ref{coef},\ref{coef1}) applies to the
whole $z$--range with $z=x/x_1$. One might try to find an effective
small--$x$
approximation. The  function is a polynomial of degree 4 in the variable
$z$. Approximating it successively  up to its $z^0, z, z^2 \ldots$ 
-contribution
one may study the effect of the individual terms. Figure~5 shows,
that a sufficient representation is      only  given taking all
contributions into account since all approximants diverge for 
$x \rightarrow 0$, i. e. an effective small-$x$ approximation
to this function does not exist.
\section{Conclusions}

\vspace{1mm}
\noindent
A           study has been performed to the slope of $F_2$ due to
twist--4 coefficient functions for the process $4 G \rightarrow 2 q$.
The present calculation was performed in time--ordered perturbation theory
and limited to $dp_{\perp}^2/p_{\perp}^4$ accuracy of the respective
Feynman diagrams. Numerical results were choosing a special ansatz for
the non--perturbative two--particle gluon density ${\bf G_2}$. The
new contributions contain anti--screening and screening terms, the latter
of which dominate for smaller values of $x$ and diminish the growth of
the slope due to the twist--2 contributions. The numerical results
show that this effect is of significant size only below $x \sim 2 \cdot
10^{-6}$ for $Q^2 \sim 5 \GeV^2$. This region is yet beyond the kinematic
range which can be probed at the $ep$ collider HERA but may be
investigated at future lepton--hadron facilities operating at larger
cms energies.

\vspace{4mm} \noindent
{\bf Acknowledgment:}~For discussions we would like to thank
W.L. van Neerven, A. Vogt and A.H.~Mueller. J.B. is thankful to
R. Devenish for financial support during the 2000 Oxford Small $x$
Meeting, where first results on this work were presented.


\newpage
\begin{center}

\mbox{\epsfig{file=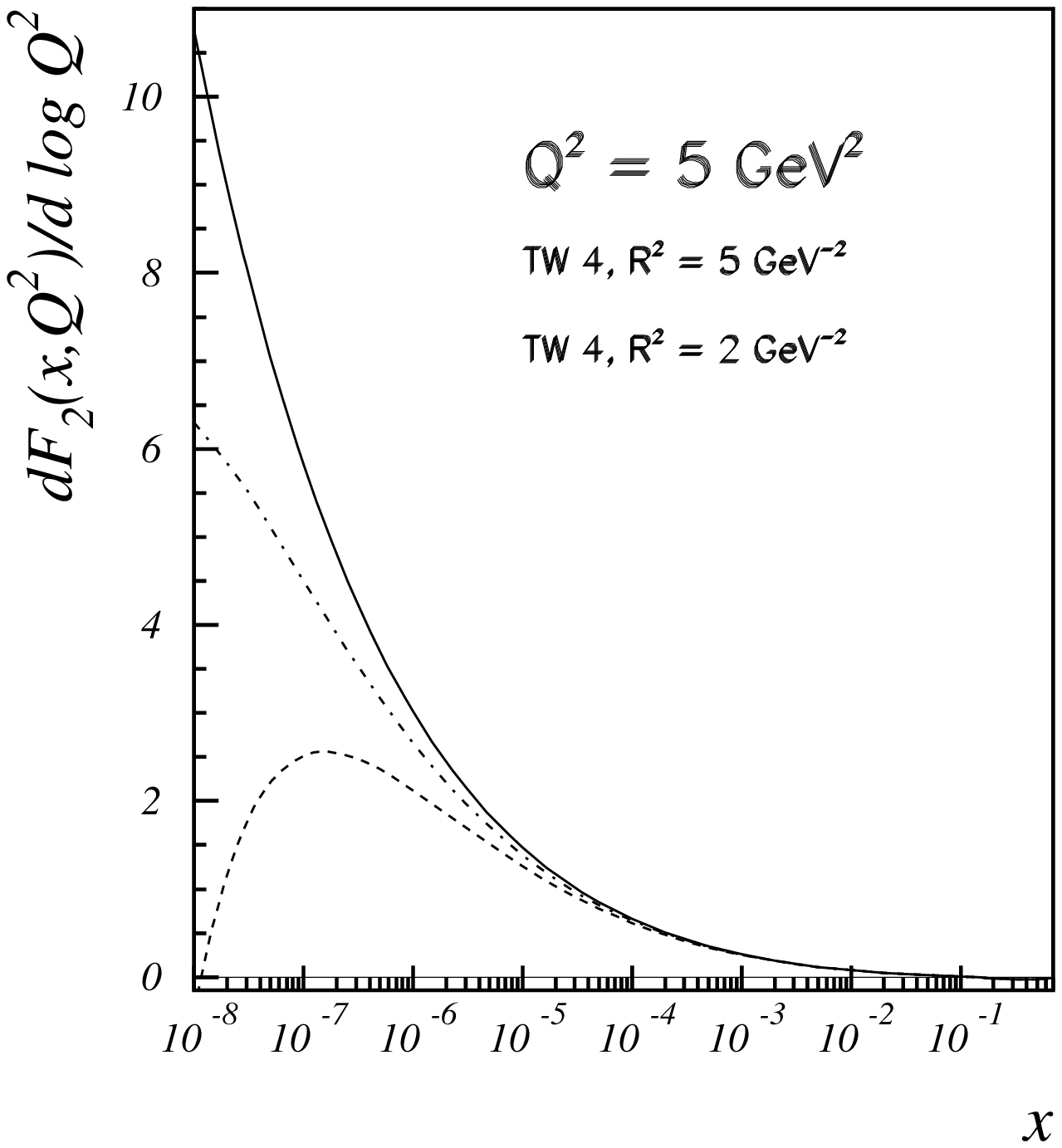,height=16cm,width=16cm}}

\vspace{2mm}
\noindent
\small
\end{center}
{\sf
Figure~4a:~The slope $dF_2(x,Q^2)/d\log Q^2$ at $Q^2 = 5 \GeV^2$.
Full line: leading order twist--2 contributions (parameterization 
Ref.~\cite{GRV98}). Dash-dotted line:  Eq.~(\ref{evol}) with twist--4
mass scale $R^2 = 5 \GeV^{-2}$, and dashed line: $R^2 = 2 \GeV^{-2}$.
\normalsize
\newpage
\begin{center}

\mbox{\epsfig{file=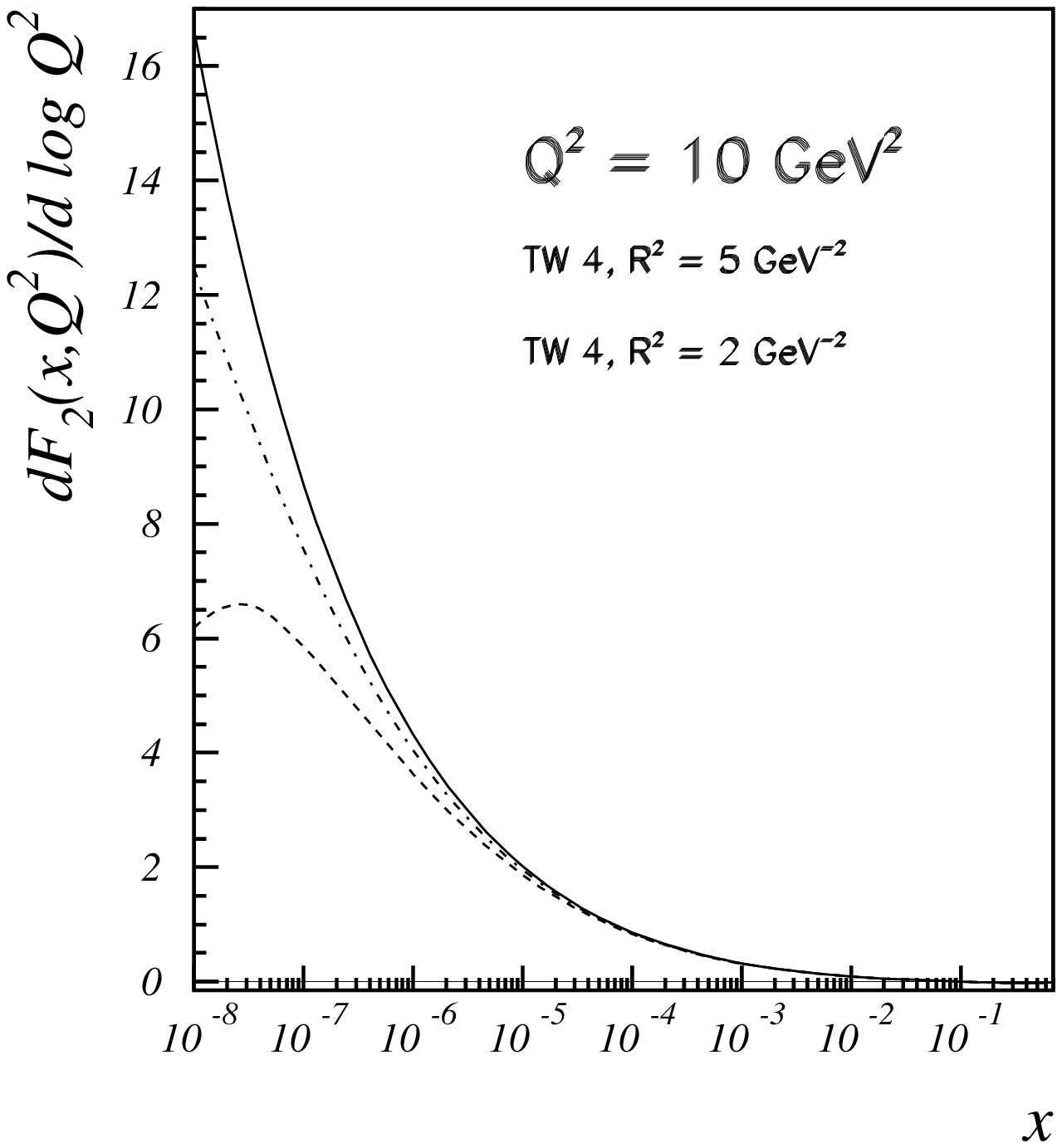,height=16cm,width=16cm}}

\vspace{2mm}
\noindent
{\sf
Figure~4b:~Same as figure~1a for $Q^2 = 10 \GeV^2$.
}
\end{center}
\normalsize
\newpage
\begin{center}

\mbox{\epsfig{file=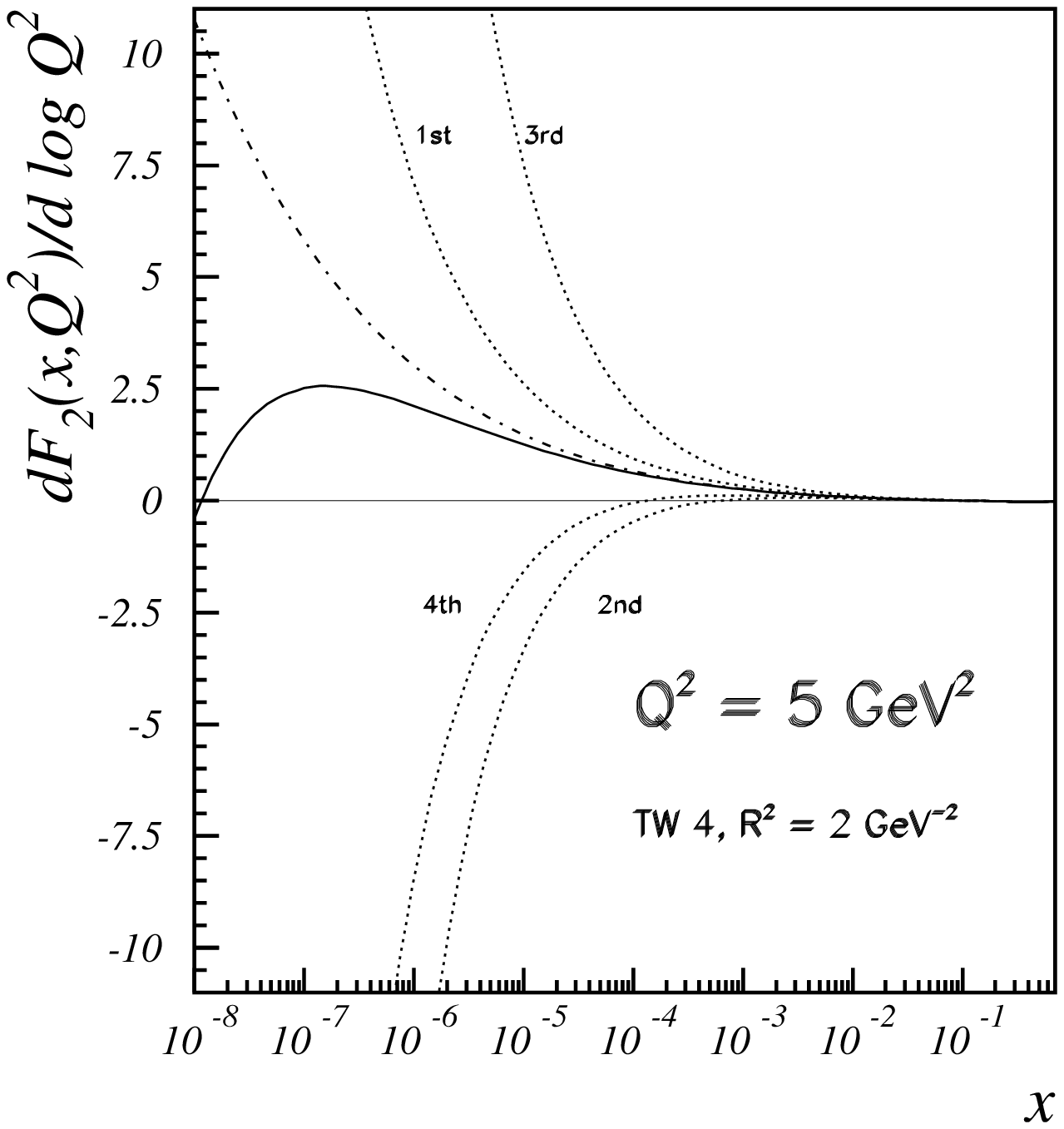,height=16cm,width=16cm}}

\vspace{2mm}
\noindent
\small
\end{center}
{\sf
Figure~5:~Comparison of the slope $dF_2(x,Q^2)/d\log Q^2$ at
$Q^2 = 5 \GeV^2$ and twist--4 mass scale $R^2 = 2 \GeV^{-2}$,
Eq.~(\ref{evol}) (full line) with the corresponding results obtained
approximating the coefficient function Eq.~(\ref{coef}) by the sequence
of contributing powers. 1st: $z^0$, 2nd: $z$ etc. (dotted lines).
Dash-dotted line: twist--2 contribution.
\normalsize
\end{document}